\documentclass{article}
\usepackage{amssymb,amsmath,amsthm}

\newcommand{\cf}{\emph{cf}}
\newcommand{\Nat}{\mathbb{N}}
\newcommand{\Int}{\mathbb{Z}}

\newcommand{\Hilbert}{\mathcal{H}}
\newcommand{\PT}{\mathcal{PT}}
\newcommand{\slim}{\mathop{\mathrm{s}\mbox{--}\lim}}

\newtheorem{Theorem}{Theorem}
\theoremstyle{remark}
\newtheorem{Remark}{Remark}
\begin{document}
%
%
%-------%
% TITLE %
%-------%
%------------------------------------------%
%------------------------------------------%
\title{\textbf{\Large
Calculation of the metric in the Hilbert space
of a $\mathcal{PT}$-symmetric model
via the spectral theorem
}}
\author{\textsc{David Krej\v{c}i\v{r}\'{\i}k}}
\date{
\footnotesize
\begin{center}
\emph{
Department of Theoretical Physics, Nuclear Physics Institute,
\\
Academy of Sciences, 250\,68 \v{R}e\v{z} near Prague, Czech Republic
\smallskip \\
\emph{E-mail:} krejcirik@ujf.cas.cz
}
\end{center}
12 July 2007}
\maketitle
\begin{abstract}
\noindent
In a previous paper~\cite{KBZ} we introduced
a very simple $\PT$-symmetric non-Hermitian Hamiltonian with real spectrum
and derived a closed formula for the metric operator
relating the problem to a Hermitian one.
In this note we propose an alternative formula
for the metric operator,
which we believe is more elegant
and whose construction
-- based on a backward use of the spectral theorem
for self-adjoint operators --
provides new insights into the nature of the model.
\end{abstract}
%
%------------------------------------------%
%------------------------------------------%
%

%---------------------%
\section{Introduction}
%---------------------%
%
Although quantum mechanic is conceptually a self-adjoint theory,
there are numbers of problems that
require the analysis of non-self-adjoint operators.
The study of resonances of self-adjoint Schr\"odinger operators
via the technique of complex scaling~\cite{CFKS}
or the derivation of the famous Landau-Zener formula
for the adiabatic transition probability between eigenstates
of a time-dependent two-level system~\cite{Joye-Kunz-Pfister_1991}
are just two examples.
However, in contrast to the well understood theory
of self-adjoint operators, the non-self-adjoint theory
can be quite different
(\cf~a nice review~\cite{Davies_2002})
and is certainly less developed.
The former is much easier to analyse because of the existence
of the spectral theorem.

Recent years brought new motivations and focused attention
to aspects of problems which attracted little attention earlier.
A strong impetus comes from
the so-called $\PT$-symmetric quantum mechanics,
where the Hamiltonian~$H$ of a system is not Hermitian
but the Schr\"odinger equation is invariant under a simultaneous
change of spatial reflection~$\mathcal{P}$ and time reversal~$\mathcal{T}$
(\cf~\cite{Bender-Boettcher_1998} for the pioneering work
and~\cite{Bender_2007} for a recent review with many references).
Here the interest consists in the fact that many of the $\PT$-symmetric
Hamiltonians possess real spectra
and that the problem can be reinterpreted as a Hermitian one
in a different Hilbert space.
Indeed, and more generally, the identification is provided
by the pseudo-Hermiticity relation
\cite{GHS,Ali1,Ali2,Ali3}:
\begin{equation}\label{pseudo.Hermiticity}
  H^* \Theta = \Theta H
\end{equation}
valid on the domain of~$H$.
Here~$\Theta$ is a bounded positive Hermitian operator,
called metric.

There have been many attempts to calculate the metric operator~$\Theta$
for the various $\PT$-symmetric models of interest
(\cf~\cite{KBZ} for the related references to which
we add the Swanson model
\cite{Swanson_2004,Geyer-Scholtz-Snyman_2004,Jones_2005}
and recent works \cite{Ali5.5,Ali6}).
Because of the complexity of the problem, however, it is not
surprising that most of the available formulae for~$\Theta$
are just approximative, usually expressed as leading terms of
perturbation series.
Moreover, the calculations are usually formal in the sense
that the boundedness of~$\Theta$ is not verified
(\cf~\cite{Kretschmer-Szymanowski_2004}
for a rigorous discussion on the importance of the boundedness).

For these reasons we decided in~\cite{KBZ}
to introduce a new one-parametric non-Hermitian $\PT$-symmetric
Hamiltonian~$H_\alpha$ with real spectrum
-- which very likely represents the simplest possible
$\PT$-symmetric model whatsoever --
and derived a formula for its metric~$\Theta_\alpha$
in a closed form and in a rigorous manner.
The method of~\cite{KBZ} relied on the fact
that the eigenfunctions of~$H_\alpha$ can be expressed
in terms of eigenfunctions of self-adjoint operators.
Using the completeness of the latter,
the metric operator was constructed by summing up
certain series of trigonometric functions.

The ultimate objective of this note is to point out
that the series determining~$\Theta_\alpha$
can be summed up alternatively -- and probably more elegantly --
by using the spectral theorem.
Moreover, we believe that the resulting formula for the metric
has a more transparent structure than that presented in~\cite{KBZ}
and might be useful for further applications
of the pseudo-Hermiticity of our model.

For the convenience of the reader
we state here a simple version
of the spectral theorem we shall use later:
\begin{Theorem}[Spectral Theorem]\label{ST}
Let~$H$ be a self-adjoint operator with compact resolvent
in a Hilbert space with inner product $(\cdot,\cdot)$,
antilinear in the first factor and linear in the second one.
Then
\begin{equation}\label{decomposition}
  f(H) = \sum_{j=0}^\infty f(E_j) \, \psi_j (\psi_j,\cdot)
\end{equation}
for any complex-valued, continuous function~$f$.
Here $\{E_j\}_{j=0}^\infty$ and $\{\psi_j\}_{j=0}^\infty$
denote respectively the set of eigenvalues
and corresponding eigenfunctions of~$H$.
\end{Theorem}

We refer to~\cite[Sec.~VI.5]{Kato}
for a proof and a more general version of the spectral theorem
when the compactness assumption is relaxed.
Similar spectral decompositions
hold also for normal operators,
but they are in general false in the non-self-adjoint theory.
Therefore it is remarkable that a modified version
of~\eqref{decomposition} with $f(E)=E^n$, $n\in\Nat$,
still holds for our non-Hermitian operator~$H_\alpha$
(\cf~\cite[Prop.~4]{KBZ} for the case $n=0$,
the other cases being a consequence).

The spectral theorem is usually used to construct
a function of a self-adjoint operator
in terms of the sum of spectral projections.
In this note we use it backwards:
we identify eigenprojections of a self-adjoint operator
and replace an infinite series by a function of the operator.

In the forthcoming Section~\ref{Sec.model} we recall
the model introduced in~\cite{KBZ}
(we refer to that reference for more details and other results).
This is followed by Section~\ref{Sec.metric}
where the alternative formula for the metric is established.

%------------------%
\section{The model}\label{Sec.model}
%------------------%
%
The simplicity of the Hamiltonian~$H_\alpha$
consists in that it acts as the free Hamiltonian
$$
  H_\alpha\psi:=-\psi''
  \qquad\mbox{in}\qquad (0,d)
  \,,
$$
while the non-Hermiticity enters uniquely through
complex Robin boundary conditions:
\begin{equation}\label{bc}
  \psi'(0)+i\alpha\psi(0)=0
  \qquad\mbox{and}\qquad
  \psi'(d)+i\alpha\psi(d)=0
  \,,
\end{equation}
where~$\alpha$ is a real constant.
It was shown in~\cite{KBZ} that~$H_\alpha$,
with the domain $D(H_\alpha)$ consisting of all functions~$\psi$
in the Sobolev space $W^{2,2}((0,d))$
such that~\eqref{bc} holds,
is an $m$-sectorial operator in~$\Hilbert:=L^2((0,d))$.
The $\PT$-symmetry of our model is reflected by the relation
$$
  H_\alpha^*=H_{-\alpha}
  \,.
$$
\begin{Remark}
A more general class of one-dimensional Schr\"odinger operators
with non-Hermitian boundary conditions of the type~\eqref{bc}  
was studied previously by Kaiser, Neidhardt and Rehberg 
in~\cite{Kaiser-Neidhardt-Rehberg_2003}. 
In their paper, motivated by the needs of semiconductor physics,
the parameter~$\alpha$ is allowed to be complex 
but its imaginary part has opposite signs on the boundary points
such that the system is dissipative.
\end{Remark}

It was also shown in~\cite{KBZ} that
the spectrum of~$H_\alpha$ is purely discrete
and given by
\begin{equation}\label{spectrum}
  \sigma(H_{\alpha})
  = \big\{\alpha^2\big\}
  \cup \big\{k_j^2\big\}_{j=1}^\infty
  \,,\qquad\mbox{where}\qquad
  k_j := j\pi/d \,.
\end{equation}
Moreover, all the eigenvalues are simple
provided
\begin{equation}\label{hypothesis}
  \alpha d/\pi \ \not\in \ \Int\!\setminus\!\{0\}
  \,.
\end{equation}
Assuming this non-degeneracy condition,
the eigenfunctions of the adjoint~$H_\alpha^*$
corresponding to the eigenvalues counted as in~\eqref{spectrum}
can be chosen as
\begin{equation}\label{phi}
  \phi_j^\alpha(x) :=
  \left\{
  \begin{aligned}
    & \chi_0^N + \rho_\alpha(x)
    & \mbox{if} & \quad j=0 \,,
    \\
    &
    \chi_j^N(x) + i \frac{\alpha}{k_j} \, \chi_j^D(x)
    &\mbox{if} & \quad j \geq 1 \,.
  \end{aligned}
  \right.
\end{equation}
Here
$$
  \rho_\alpha(x) :=
  \frac{\exp(i\alpha x) - 1}{\sqrt{d}}
$$
and $\{\chi_j^N\}_{j=0}^\infty$,
respectively $\{\chi_j^D\}_{j=1}^\infty$,
denotes the complete orthonormal family
of the eigenfunctions of the Neumann Laplacian $-\Delta_N$,
respectively Dirichlet Laplacian $-\Delta_D$,
in~$\Hilbert$:
\begin{equation*}
  \chi_j^N(x) :=
  \begin{cases}
    \sqrt{1/d}
    & \mbox{if}\quad j=0 \,,
    \\
    \sqrt{2/d} \, \cos(k_j x)
    & \mbox{if}\quad j \geq 1 \,,
  \end{cases}
  \qquad
  \chi_j^D(x) := \sqrt{2/d} \, \sin(k_j x)
  \,.
\end{equation*}
Note that $-\Delta_N=H_0$
and that the spectrum of~$-\Delta_D$
is equal to $\{k_j^2\}_{j=1}^\infty$.

%----------------------------------%
\section{Calculation of the metric}\label{Sec.metric}
%----------------------------------%
%
Still under the hypothesis~\eqref{hypothesis},
it was demonstrated in~\cite{KBZ} that the operator
\begin{equation}\label{metric}
  \Theta_\alpha
  := \sum_{j=0}^\infty \phi_j^\alpha (\phi_j^\alpha,\cdot)
  \equiv \slim_{m\to\infty}
  \sum_{j=0}^m \phi_j^\alpha (\phi_j^\alpha,\cdot)
\end{equation}
is bounded, symmetric, positive
and satisfying~(\ref{pseudo.Hermiticity}) with~$H_\alpha$.
Here $(\cdot,\cdot)$ denotes the inner product in~$\Hilbert$,
antilinear in the first factor and linear in the second one.
Furthermore, a closed integral-type formula
for the operator was derived by using known results
about the sum of trigonometric functions.

Now we propose an alternative way
how to sum up the infinite series in~\eqref{metric}.
First we write~$\Theta_\alpha$ as
$$
  \Theta_\alpha
  = P_0^\alpha
  + \Theta^{(0)}
  + \alpha \, \Theta^{(1)}
  + \alpha^2 \, \Theta^{(2)}
$$
with
\begin{align*}
  P_0^\alpha
  &:= \phi_0^\alpha (\phi_0^\alpha,\cdot)
  = P_0^N + \chi_0^N (\rho_\alpha,\cdot)
  + \rho_\alpha (\chi_0^N,\cdot)
  + \rho_\alpha (\rho_\alpha,\cdot)
  \,,
  \\
  \Theta^{(0)}
  &:= \sum_{j=1}^\infty \chi_j^N (\chi_j^N,\cdot)
  = I - P_0^N
  \,,
  \\
  \Theta^{(1)}
  &:= \sum_{j=1}^\infty
  \Big(
  -i k_j^{-1} \, \chi_j^N (\chi_j^D,\cdot)
  + i k_j^{-1} \, \chi_j^D (\chi_j^N,\cdot)
  \Big)
  \,,
  \\
  \Theta^{(2)}
  &:= \sum_{j=1}^\infty
  k_j^{-2} \, \chi_j^D (\chi_j^D,\cdot)
  = (-\Delta_D)^{-1}
  \,,
\end{align*}
where $P_0^N:=\chi_0^N (\chi_0^N,\cdot)=P_0^0$
and~$I$ denotes the identity operator in~$\Hilbert$.
The equalities in the second and fourth lines
follow directly by Theorem~\ref{ST}
applied to $-\Delta_N$ and $-\Delta_D$, respectively.
In order to use the spectral theorem in~$\Theta^{(1)}$ as well,
we introduce a ``momentum'' operator~$p$ in~$\Hilbert$ by
\begin{equation}\label{momentum}
  p\psi:=-i\psi'
  \,,
  \qquad
  D(p):=W_0^{1,2}((0,d))
  \,.
\end{equation}
The adjoint operator~$p^*$ acts in the same way
but has a larger domain, $D(p^*)=W^{1,2}((0,d))$.
Since $\chi_j^D$ and $\chi_j^N$ belong to
$D(p)$ and $D(p^*)$, respectively, we have
$p\chi_j^D=-ik_j\chi_j^N$ and $p^*\chi_j^N=ik_j\chi_j^D$.
Consequently, Theorem~\ref{ST} yields
\begin{align*}
  \Theta^{(1)} &=
  p \sum_{j=1}^\infty
  k_j^{-2} \, \chi_n^D (\chi_n^D,\cdot)
  + p^* \sum_{j=1}^\infty
  k_j^{-2} \, \chi_n^N (\chi_n^N,\cdot)
  \\
  &= p\,(-\Delta_D)^{-1} + p^* (-\Delta_N^\bot)^{-1}
  \,,
\end{align*}
where $-\Delta_N^\bot := (I-P^N)(-\Delta_N)(I-P^N)$.
Notice that the ``interchange of summation and differentiation''
in the first equality is justified just by the definition
of the sum in~\eqref{metric} and
the distributional derivative in~\eqref{momentum}.

Summing up, we get
\begin{Theorem}
The linear operator $\Theta_\alpha$ in~$\Hilbert$ defined by
\begin{equation}\label{result}
  \Theta_\alpha =
  I + P_0^\alpha - P_0^N
  + \alpha p\,(-\Delta_D)^{-1}
  + \alpha p^* (-\Delta_N^\bot)^{-1}
  + \alpha^2 (-\Delta_D)^{-1}
\end{equation}
is bounded, symmetric, non-negative
and satisfies~\eqref{pseudo.Hermiticity} with~$H_\alpha$.
Furthermore, $\Theta_\alpha$ is positive
if the condition~\eqref{hypothesis} holds true.
\end{Theorem}

Note that the metric~$\Theta_\alpha$ tends to~$I$ as $\alpha \to 0$,
which is expected due to the fact that~$H_0$
coincides with the self-adjoint operator~$-\Delta_N$.

\begin{Remark}
Formula~\eqref{result} can be written exclusively
in terms of the operators~$p$ and~$p^*$ by employing the identities
$-\Delta_D=p^*p$ and $-\Delta_N=pp^*$.
Note also that the resolvent $(-\Delta_D)^{-1}$
and the reduced resolvent $(-\Delta_N^\bot)^{-1}$
are integral operators
with explicit and extremely simple kernels
(\cf~\cite[Ex.~III.6.21]{Kato}).
\end{Remark}
%

%---------------------------%
\subsection*{Acknowledgment}
%---------------------------%
I am grateful to Miloslav Znojil for many valuable discussions.
The work has been supported by FCT, Portugal,
through the grant SFRH/BPD/11457/2002,
and by the Czech Academy of Sciences and its Grant Agency
within the projects IRP AV0Z10480505 and A100480501,
and by the project LC06002 of the Ministry of Education,
Youth and Sports of the Czech Republic.

%--------------%
% BIBLIOGRAPHY %
%--------------%
%
{\small
%\bibliography{bib}
%\bibliographystyle{amsplain}

\providecommand{\bysame}{\leavevmode\hbox to3em{\hrulefill}\thinspace}
\providecommand{\MR}{\relax\ifhmode\unskip\space\fi MR }
\providecommand{\MRhref}[2]{%
  \href{http://www.ams.org/mathscinet-getitem?mr=#1}{#2}
}
\providecommand{\href}[2]{#2}

}
\end{document}